\documentclass[pre,twocolumn,showpacs,superscriptaddress]{revtex4}
\usepackage{amsmath,amssymb}
\usepackage{graphics,graphicx}
\usepackage{dcolumn,bm}
\usepackage{psfrag}
\usepackage[utf8x]{inputenc}
\newcommand{\icf}
{\affiliation{Instituto de Ciencias F\'isicas,Universidad Nacional Aut\'onoma de M\'exico, Cuernavaca 62210, M\'exico }}

\begin{document}

\title{Ballistic annihilation with superimposed diffusion in one dimension}

\author{Soham Biswas}%
\email{soham@fis.unam.mx}
\icf 

\author{Hern\'an Larralde}%
\email{hernan@fis.unam.mx}
\icf

\author{Francois Leyvraz}%
\email{leyvraz@fis.unam.mx}  
\icf

\begin{abstract}
We consider a one-dimensional system with particles having either
positive or negative velocity, which annihilate on contact. To the
ballistic motion of the particle, a diffusion is superimposed.  The
annihilation may represent a reaction in which the two particles yield
an inert species. This model has been the object of previous work, in
which it was shown that the particle concentration decays faster than
either the purely ballistic or the purely diffusive case.  We report
on previously unnoticed behaviour for large times, when only one of
the two species remains and also unravel the underlying fractal
structure present in the system.  We also consider in detail the case
in which the initial concentration of right-going particles is
$1/2+\varepsilon$, with $\varepsilon\neq0$. It is shown that a
remarkably rich behaviour arises, in which two crossover times are
observed as $\varepsilon\to0$.
\end{abstract}

\pacs{02.50.-r, 82.20.Db, 05.20.Dd, 89.75.Da}
\maketitle

\section{Introduction}
\label{sec:intro}
The deviations from mean-field theory caused by fluctuations have been
the object of considerable research \cite{review}.  Thus, when reactants
diffuse and react on contact, they create spatial correlations which
invalidate the usual concept of mean-field theory, and indeed,
invalidate the applicability of rate equations, which are normally
viewed as fundamental in chemical kinetics. A very simple example of
this is one-species annihilation (or aggregation) in which one species
reacts with itself via a bimolecular reaction
\begin{equation}
A+A\mathop{\longrightarrow}_K B
\label{eq:1new}
\end{equation}
where $B$ represents an inert species, which we disregard in the
following. The rate equation for such a process is given by
\begin{equation}
\dot c_A=-Kc_A^2,
\label{eq:2new}
\end{equation}
and the decay for large times is therefore $1/t$. However, as is
well-known, see for example \cite{sp}, if we consider a model of point
random walkers on the line, which annihilate whenever two walkers
alight on the same site, then the decay is in fact given by
$(Dt)^{-1/2}$.  At a deeper level, the amplitude of the $1/t$ decay
determined by (\ref{eq:2new}) depends on the initial concentration,
whereas the amplitude of the $1/\sqrt t$ decay does not.

These, and many similar results have been discussed extensively. For a
review, see for example the book \cite{review_book} by Redner and
Krapivsky. In the following, we shall consider a variation on the
model described by (\ref{eq:1new}), in which the transport has both a
ballistic component, in which the particles can have one of two
velocities $v$ and $-v$, and a diffusive component.  The purely
ballistic model was initially discussed by Elskens and Frisch in
\cite{bal}. Similarly to the diffusive model discussed in \cite{sp},
the decay of concentration goes as $t^{-1/2}$, but for quite different
reasons. As was then pointed out in \cite{ballistic, baldif},
combining both diffusion and ballistic motion leads, in one dimension,
to a decay that is more rapid than the one brought about by either mechanism.

The kinetics of particle annihilation, $A + A \rightarrow \emptyset$
has attracted considerable attention from the scientific community
from another completely different point of view. The dynamics of
particle annihilation with $A + A \rightarrow \emptyset$, have a
direct one to one correspondence with the kinetic Ising model in one
dimension. For example, the zero temperature Glauber dynamics in one
dimensional Ising spin system with nearest neighbour interaction can
be mapped to the $A + A \rightarrow \emptyset$ system where the $A$
are pure random walkers \cite{Francois}. On the other hand, the binary
opinion dynamic models in one dimension (which can be directly mapped
to one dimensional Ising spin system), have a much more complicated
walker picture, where the $A$ walkers are not purely random. There are
also binary opinion dynamic models, where $A$ walkers perform a
complicated ballistic motion. For example, in the dynamics, introduced
in \cite{war}, the boundaries of the domains (which could be construed
as walkers) move ballistically unless they meet the boundary of some
other domain. Once two boundaries meet, the annihilation of the
boundaries is more involved than the simple $A + A \rightarrow
\emptyset$. In another example of a binary opinion dynamics model
introduced in \cite{opn}, the walkers in the corresponding walker
picture have ballistic motion, in this case, the $A$ walkers always
move ballistically in the direction of their nearest walker and
annihilate upon meeting, following $A + A \rightarrow \emptyset$. 
This leads to complex changes in the direction of motion of the
walkers, somewhat akin to a random walk \cite{opn}. Understanding the
dynamics of simple ballistic annihilation process with superimposed
diffusion can thus also be of help in understanding these complex
dynamics.

The simple model of ballistic annihilation discussed in this paper has
a quite rich structure. Introducing a corresponding binary opinion
dynamics model could be a motivation and direction for the future
work.

Here we study the model of ballistic annihilation with superimposed
diffusion in greater detail than was previously done, showing in
particular a remarkably rich structure when the number of left- and
right-going particles is allowed to be different. In particular, we
study the nature of the crossovers appearing as these numbers come
close to each other.

\section{The model}
\label{model}
\subsection{Description of the model}

 Let us first describe in some detail the model we study in this
  paper. The system we consider consists of point particles on a one
  dimensional lattice. Each particle moves, always, either to the left
  or to the right, that is, each particle has positive or negative
  velocity.  Initially there is a fraction ($1/2 + \varepsilon$) of
  particles with positive velocity and $(1/2 -\varepsilon)$ of
  particles with negative velocity (with $-1/2\leq\varepsilon\leq1/2$)
  randomly distributed on the lattice. Whenever a particle lands on a
  site in which another particle is found, both are removed from the
  system.  This is therefore a model of the so-called one species
  annihilation type ($A + A \rightarrow \emptyset$). It is of central
  importance that the particles are moved asynchronously, that is, at
  each time step, one particle is chosen at random and moved in the
  direction corresponding to the velocity of that particle.

Note that the asynchronous updating rule leads to the possibility that
particles having the same velocity can react, since the random choice
of the particle leads to an effective diffusive motion of the
particles with respect to their neighbouring particles having the same
speed, whereas in a perfectly synchronous update, these particles'
distances would remain fixed.  It is largely the consequences of this
effect we shall explore in this paper.  This model is equivalent to a
model in which a fraction $1/2+\varepsilon$ of particles perform a
random walk with bias $v$, whereas the others have a bias $-v$. We
shall generally work with a bias equal to $v$ in order to keep
dimensions explicit. 

For $\varepsilon =0 $, the number of positive
and negative velocity particles are equal, on average, at the
beginning. Taking $\varepsilon \neq 0 $ introduces inequality in the
numbers of positive and negative velocity particles, and $ \varepsilon
= \pm 1/2 $ means there is only one kind of particles.

The synchronous update version of this system has been studied
by Elskens and Frisch \cite{bal}, who showed that for $\varepsilon=0$,
the concentration of particles decays as $t^{-1/2}$. Here we study the
following variant: instead of letting the particles move ballistically
in continuous time, we discretize time and choose, at each time step,
a particle at random and move it to the right if it has positive
velocity, and to the left otherwise. In other words, we use an
asynchronous updating rather than a synchronous one. This introduces a
diffusion, and thereby the possibility for two particles having the
same velocity to annihilate with each other. This apparently minor
change affects the system profoundly, as already noted by
\cite{ballistic,baldif} in the case $\varepsilon=0$. Here we extend
the study to the general case, and also display some non-trivial large
time behaviour in the case $\varepsilon=0$ which had previously gone
unreported. Finally, we propose a mechanism, first studied by Alemany \cite{alemany},
for modifying the decay exponent of diffusive annihilation kinetics in
one dimension, and show that it is indeed operative in this particular
system.

\begin{figure}[ht]
\includegraphics[width=5cm,angle=270]{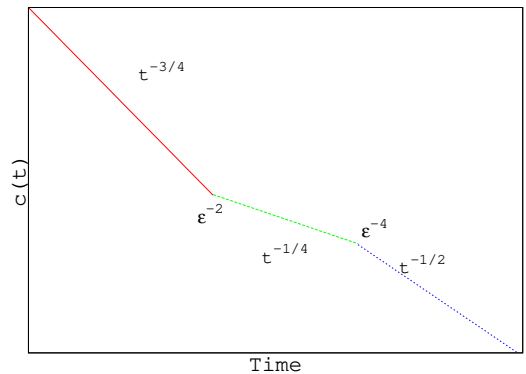}
\caption{(Color Online) A schematic figure with two crossovers for the decay of
  $c(t)$, the concentration of total number of particles at time
  $t$. }
\label{sch_cross}
\end{figure}

Let us briefly summarize our results for the general case
$\varepsilon>0$. We find three different regimes, of which the easiest
is certainly the last: at very large times, all particles with
negative velocities have disappeared. Additionally, all the spatial
correlations their presence might have induced --- we shall see that
such correlations can in fact arise --- have also disappeared. We thus
have a system consisting solely of right-moving particles moving
ballistically, with a superimposed diffusion. This is equivalent to
pure diffusive dynamics, so that the asymptotic decay goes as
$(Dt)^{-1/2}$. This is the third stage of the system's evolution. If
$\varepsilon\ll1$, there are two other stages: the first one is the
one in which we may neglect the difference in concentration between
left- and right-going particles. In this regime, the usual decay
exponent reported and analysed by ben-Naim, Redner and Krapivsky
\cite{baldif} applies and the concentration decays as $t^{-3/4}$. This
stage ends when all left-going particles have disappeared. This
happens, as we shall see, at a time $t_1(\varepsilon)$ of order
$\varepsilon^{-2}$ [Fig \ref{sch_cross}].  At this point, a second
stage sets in, characterised by a decay with $t^{-1/4}$ leading term
[Fig \ref{sch_cross}]. This is due to the fact that the surviving
right going particles are far from being uniformly distributed, which
leads to an anomalous decay, as pointed out by Alemany
\cite{alemany}. The third stage, when these correlations finally
disappear, sets in at a time $t_2(\varepsilon)$ that scales as
$\varepsilon^{-4}$ [Fig \ref{sch_cross}].

If we start with $N$ particles, then one finds that at
$t_1(\varepsilon)$ there remain $N\varepsilon^{3/2}$ particles,
whereas at $t_2(\varepsilon)$ there only remain
$N\varepsilon^{5/2}$. This brings out some numerically challenging
features of the model: if we wish to have a clean separation of time
scales, we need at least $\varepsilon\sim10^{-2}$, and if we wish to
have, say, a hundred particles at $t_2(\varepsilon)$ (in order to be
able to observe the final exponent reliably), we would need to start
out with $N\sim10^7$, which is altogether unrealistic. We shall
therefore rely on various simulations with different values of the
parameters to bring out the various features of the system.

\subsection{Quantities Calculated}
\label{subsec:quant}
We have studied the following quantities in the present work. 

\begin{enumerate}
\item Decay of the concentration of total number particles $c(t)$ with
  time, starting from concentration $c(0)=1$, which means that all sites are
  occupied initially.

\item Decay of the concentration of excess number of particles
  $c_{ex}(t)$ with time, which is defined as the absolute difference
  between positive and negative velocity particles. For $\varepsilon
  =0 $, $c_{ex}(0) \simeq 1/\sqrt{N}$ (up to an $O(1)$ factor) where $N$ is the initial number of
  particles.

\item Change of the domain size of positive and negative velocity
  particles with time.

\item The distribution of inter-particle distances between two
  neighbouring particles (independent of their velocities) at late
  time. This distribution is also studied for same velocity particles
  and different velocity particles.

\item The persistence probability $P(t)$: This is the probability that
  a site remains unvisited by any of the walkers $A$ up to time
  $t>0$. When the walkers perform pure random walk, the persistence
  probability $P(t)$ shows a power law decay given by $P (t) \sim
  t^{-\theta}$ , where $\theta$ is the persistence exponent and is
  unrelated to any other known static or dynamic exponents \cite{persistence, per}. 
\end{enumerate}

We have studied the dynamics starting from initially $N$ 
particles, with $10^4 \leq N \leq 8\cdot10^4$. The results are
averaged over $2000$ to $2500$ configurations. Periodic boundary
conditions have been used.

\section{Theory }
\label{sec:theory}

For $\varepsilon =\pm 1/2 $ only one kind of ballistic particles
(either $+v$ or $-v$ velocity particles) exist in the system. Due to
the random update rule for the simulation, by which diffusion is
incorporated, the relative motion of the particles is actually
diffusive in this case. The relative diffusion constant
$D_{eff}=1$ is the same as if the particles performed symmetric random
walks (see Appendix \ref{diffconst} for details). Naturally, in this
situation, the concentration $c(t)$ will decay as $t^{-1/2}$.

If $\varepsilon \neq \pm 1/2 $, for any typical configuration, the
number of $+v$ and $-v$ velocity particles will not be equal (this is
clear for $\varepsilon \neq 0$, but also when $\varepsilon = 0$, due
to statistical fluctuations).  Due to this inequality in numbers, the
system will go from a regime of ballistic annihilation with
superimposed diffusion, to a long time regime of pure diffusion when
only one kind of particle is present. In the following, we start by
considering the regime in which no significant difference in the
numbers of left- and right-going particles exists $\varepsilon =
0$. In this case, as has been shown in \cite{ballistic,baldif}, the
particle number decays as $t^{-3/4}$. In the following subsection, we
shall re-derive this result for the sake of keeping the paper
self-contained. Afterwards, we proceed to analyse the case of finite
systems, in which the particles eventually all have one given velocity
due to the effect of statistical fluctuations. Finally we analyse the
case of $\varepsilon\neq0$.

\subsection{The infinite system with $\varepsilon=0$: dimensional analysis}
\label{subsec:infinite}

In the following, we analyze the system for $\varepsilon=0$ using
dimensional considerations.  The various parameters involved are
$c(0)$, $D$, $t$ and $v$. These can be combined in two dimensionless
parameters:
\begin{subequations}
\begin{eqnarray}
x&=&\frac{v}{Dc(0)}
\label{eq:dimlessa}\\
\tau&=&\frac{v^2t}{D}
\label{eq:dimlessb}
\end{eqnarray}
\label{eq:dimless}
\end{subequations}
The former is the ratio of the time required for nearest neighbours to
cross the average distance between them if they move ballistically, to
the time required to cross the same distance with diffusive
dynamics. It thus states whether diffusion or drift dominates the
short-time dynamics. $\tau$ is a dimensionless time, which separates
the regime in which diffusion dominates, $\tau\ll1$, from that in
which drift dominates, $\tau\gg1$

The concentration of particles at time $t$ can thus be written in
terms of the adimensional quantities $x$ and $\tau$ as follows:
\begin{equation}
c(t) \simeq c(0)\Phi\!\left(
x, \tau
\right)
\label{dim}
\end{equation}

If $x\ll1$, then $c(t)$ should not depend on $v$ for initial times,
since the process is (in a first approximation) purely diffusive,
leading to a $(Dt)^{-1/2}$ behaviour.  Therefore, for $x\ll1$,
  using the expression of $x$ and $\tau$, given by equation
  (\ref{eq:dimlessa}) and (\ref{eq:dimlessb})  we get
\begin{equation}
\Phi(x, \tau)=\frac{x}{\sqrt\tau}
\label{weakbias}
\end{equation}
This is only valid up to $\tau\sim1$, since, as we have seen above,
this is the crossover time between drift and diffusion.

On the other hand, for $x\gg1$, the behaviour of $c(t)$ is purely
ballistic for moderate $\tau$.  It is thus independent of $D$ and
given by $\sqrt{c(0)/vt}$, as shown in \cite{bal}. Therefore, for
$x\gg1$, again using the equations (\ref{eq:dimlessa})
  and(\ref{eq:dimlessb}) we find
\begin{equation}
\Phi(x, \tau)=\sqrt{\frac{x}{\tau}}
\label{strongbias}
\end{equation}
Let us now determine the time $\tau$ up to which this is valid. The
qualitatively new feature introduced by diffusion is the possibility
for particles of the same velocity to annihilate via diffusion. But
this happens on a time scale $1/(Dc(0)^2)$, which, in terms of the
adimensional quantities, is $\tau\sim x^2$. We see, therefore, that
the approximation (\ref{strongbias}) should hold up to $\tau\sim x^2$

The previous equations (\ref{weakbias}) and (\ref{strongbias}) only
hold for short times.  For the former, this is because the influence
of drift is eventually felt, whereas for the latter, it is due to the
diffusive annihilation of particles with like velocity. If we now
describe the large $\tau$ behaviour for the full dynamics, by
\begin{equation}
\Phi(x, \tau)\simeq x^\alpha\tau^\beta
\label{asympt}
\end{equation}
then we obtain two conditions on $\alpha$ and $\beta$ as follows: when
$\tau\sim 1$ and $x\ll1$, we may apply both (\ref{weakbias}) and
(\ref{asympt}).  This leads to
\begin{equation}
x\sim x^\alpha
\end{equation}
implying $\alpha=1$. Similarly, if we consider the case in which
$x\gg1$ and $\tau\sim x^2$, we may apply (\ref{strongbias}) as well as
(\ref{asympt}), and are thereby led to
\begin{equation}
\sqrt{\frac{1}{x}}\sim x^{\alpha+2\beta}=x^{1+2\beta},
\end{equation}
from which follows 
\begin{equation}
\beta=-\frac{3}{4}.
\end{equation}

We thus obtain
\begin{equation}
\Phi(x, \tau)\simeq x\tau^{-3/4}.
\label{final}
\end{equation}
This finally yields, for the concentration $c(t)$ at
large times:
\begin{equation}
c(t)\simeq v^{-1/2}D^{-1/4}t^{-3/4}.
\label{powerz1}
\end{equation}
This dimensional analysis parallels that made in \cite{ballistic,
  baldif} and is presented for the sake of keeping the paper
self-contained.

Note that the above derivation contains a weak point: it is assumed in
(\ref{asympt}) that the $x$ dependence of the prefactor of
$\tau^\beta$ is always the same power $\alpha$, both in the $x\ll1$
and in the $x\gg1$ regime. This is, in itself, not obvious, and we
shall see in later sections a derivation of the same result, free from
this objectionable feature.

\subsection{The case of finite systems}
\label{subsec:finite}

Let us first describe the behaviour of the set of surviving particles in 
ballistic annihilation with synchronous updating. Here we follow
\cite{bal}.
 
Let the initial condition of a system undergoing ballistic
annihilation be given by the numbers $\sigma_k$, where $k$ runs from
$0$ to $L$, where $L$ is the length of the system, and $\sigma_k$ can
take three possible values: $\sigma_k=\pm1$ means that site $k$ is
initially occupied by a particle having velocity $\sigma_k$, whereas
$\sigma_k=0$ means that site $k$ is initially unoccupied. Assume the
$\sigma_k=1$ represent the $+v$ velocity particles and
the $\sigma_k=-1$ represent the $-v$ velocity particles.  Under
these circumstances, once the initial condition is set, something we
shall assume to have been done at random, then the fate of each
particle is uniquely determined. Indeed, each particle has a unique
annihilation partner, or else survives indefinitely. If $\sigma_k=1$,
the unique annihilation partner is initially at position $\pi_+(k)$,
defined by the following condition: let ${\cal A}_k$ be defined as the
following set:
\begin{equation}
{\cal A}_k:=\left\{m\in{\mathbb N}:
\sum_{r=k}^{k+m}\sigma_r=0.
\right\}
\label{eq:1}
\end{equation}
Then $\pi_+(k)$ is the smallest element of ${\cal A}_k$. If ${\cal
  A}_k=\emptyset$, then the particle initially at $k$ survives
indefinitely and $\pi_+(k)=\infty$ (a wholly similar definition works
if $\sigma_k=-1$, in which case we would say the partner is at initial
position $\pi_-(k)$).

Given the initial condition, each particle survives until it
encounters its reaction partner. The collision time is thus
\begin{equation}
\tau(k)=\frac{\pi_+(k)}{2},
\label{eq:2}
\end{equation}
where we have considered the positive velocity particles. 

We now determine the structure of the set $\Sigma_t$ defined as
\begin{equation}
\Sigma_t=\left\{
k:\tau_+(k)>t
\right\}
\label{eq:3}
\end{equation}
Let us consider a pair of particles with positive velocity separated
by a distance $k$. Without loss of generality we may assume that we
have two particles, one at $0$ and one at $k>0$. The particle at $k$
we call the {\it leader}, the one at zero, the {\it follower}. For
both to belong to $\Sigma_t$, the following conditions are necessary:
\begin{enumerate}
\item ${\cal A}_0$ should not have any element $r\leq t$.
\item ${\cal A}_k$ should not have any element $s\leq t$
\end{enumerate}
If $k>t$, then the two intervals $[0,t]$ and $[k,k+t]$ are
disjoint. The probability of both sites belonging to $\Sigma_t$ is
thus simply the product of either site belonging to $\Sigma_t$ and no
dependence on $k$ exists.  Note that this is so whether or not
$\varepsilon=0$ in the initial condition.

Let us now consider the opposite case. In this case, the leader
``clears the way'' for the follower: since ${\cal A}_k$ has no element
$s\leq t$, it follows that ${\cal A}_0$ cannot contain any elements in
the interval $[k,k+t]$. Thus, for the follower to belong to
$\Sigma_t$, it is sufficient that there be no elements in ${\cal A}_0$
belonging to the interval $[0,k]$. The probabilities that the leader
and the follower both belong to $\Sigma_t$ are thus again a product,
but this time of the probability that for all $r<k$ we have
\begin{equation}
\sum_{m=0}^r\sigma_m>0
\label{eq:4}
\end{equation}
In other words, this is the probability $p_0(k)$ that a random walk,
which starts at the origin and takes a step to the right does not
return to the origin before time $k$. The probability for this, as is
well known \cite{Weiss}, scales as $k^{-1/2}$ for $k\gg1$, if the walk
is symmetric, which corresponds, in our case, to an initial condition
with $\varepsilon=0$. Thus, if $\varepsilon=0$, the probability of
having two particles separated by a distance $k<t$ both surviving a
time $t$ is of the order of $k^{-1/2}$. In other words, this
description is compatible with the set $\Sigma_t$ forming a fractal
set---below the cutoff value $t$---with fractal dimension $1/2$.  The
correlation function for $k\gg1$ and $t\gg1$ but $k<t$ is of order
$(kt)^{-1/2}$. 

\begin{figure}[ht]
\includegraphics[width=8.6cm,angle=0]{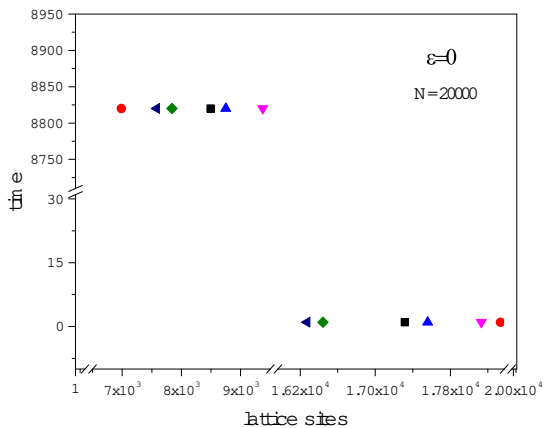}
\caption{(Color Online) Initial and very late time position of the particles which
  have survived for long enough time. The picture suggests that the
  long surviving particles are fractally distributed over the lattice
  from the beginning. This plot is generated starting from initially
  $N=2\cdot10^4$ particles. Note that the number of final particles is
  quite small---namely six---in order to claim that these lie on a
  fractal. We may nevertheless view this as evidence for the fact
  that, in the limit of infinitely many initial particles, the
  surviving particles would indeed lie on a fractal.}
\label{frac_p}
\end{figure}

The initial and very late time positions of the surviving particles as
shown the schematic figure \ref{frac_p}, form a number of clusters
(of two or more particles) and the clusters are well separated from
each other. This indicates that the long surviving particles are on a
fractal from the beginning.

In the case with no diffusion, we therefore see that $\Sigma_t$ is a
fractal with lower cutoff at length $1$ (lattice spacing) and at upper
cutoff $t$ with fractal dimension $1/2$. This was the case in which we
have synchronous updating. If instead we have asynchronous dynamics,
the annihilation of similar particles eliminates all particles with a
distance less than $\sqrt t$.  The corresponding set of surviving
particles then becomes a fractal of dimension $1/2$ with lower cutoff
$\sqrt t$ and upper cutoff $t$, leading to the fact that the set has
$t^{1/4}$ elements in each domain of size $t$, thereby leading to a
concentration $c(t)$ of $t^{-3/4}$.

We therefore see that the particles surviving at the time $t_1$, at
which only one species survives, also lie on a fractal of dimension
$d_f=1/2$.

The fact that $\Sigma_{t_1}$ is a fractal further implies, as shown by
Alemany \cite{alemany}, that the decay of a purely diffusive reaction
starting from such an initial condition is not given by $t^{-1/2}$,
but by $t^{-d_f/2}$, where $d_f$ is the fractal dimension of the
initial condition. In the particular case which concerns us here,
since $d_f=1/2$, we have a decay law of $t^{-1/4}$. This is seen in
qualitative terms as follows: the probability that a given particle
survives for a time $t$ is negligible, if this particle is both
followed and preceded by a particle significantly closer than
$\sqrt{Dt}$. We therefore require, for the particle to survive, that
one of either neighbours be further from the central particle than
$\sqrt{Dt}$, which has the probability $(Dt)^{-d_f/2}$. In our case,
this means that the final decay, after the annihilation of all
minority particles, goes as $t^{-1/4}$.

Let us look at this decay law in greater detail. When the particles
are distributed on a fractal of dimension $d_f$ ($0<d_f<1$), the
probability distribution function for initial inter-particle distances
for two nearest particles will be $P(x)$,
\begin{equation}
 P(x)=\dfrac{1}{\zeta(\lambda)}\sum_{l=1}^{\infty} l^{-\lambda}\delta(x-l)
\label{indist}
\end{equation}
where $\lambda = d_f +1$. The mean distance between nearest-neighbour
particles, diverges for $ 0 < d_f < 1 $ or for $ 1 < \lambda < 2 $. We
have obtained the decay law considering this discrete distribution
given by equation (\ref{indist}), following the formalism developed by
Alemany \cite{alemany}. The number of particles $n(t)$ at time $t$
(normalised by initial number of particles $n(0)$) will be
\begin{widetext}
\begin{equation}
 n(t)=-\dfrac{\Gamma(1-\lambda)}{2\zeta(\lambda)\Gamma(\frac{3-\lambda}{2})} \tau^{-\frac{\lambda -1}{2}} + 
 \dfrac{\Gamma(1-\lambda)}{4(1-\lambda) [\zeta(\lambda)]^2}~\tau^{-(\lambda-1)} 
 + \dfrac{\zeta(\lambda -1)}{2\zeta(\lambda)\sqrt{\pi}}~\tau^{-1/2} + O(\tau^{-\lambda/2})~~~~~~
\label{nwfrct}
\end{equation}
\end{widetext}
where $\tau=D_{eff}~t$. See Appendix \ref{fractal} for details. For
the fractal dimension $d_f=1/2$, that is $\lambda=1.5$, the leading
term will be $t^{-1/4}$ and the coefficient of this leading term is
positive, as $\Gamma(1-\lambda)$ is {\em negative\/} for all $\lambda
>1 $. The first correction to scaling will involve both the second and
the third term of (\ref{nwfrct}), thereby leading to
\begin{widetext}
\begin{eqnarray}
n(t)&=&\dfrac{\sqrt\pi}{\zeta(3/2)\Gamma(3/4)}\tau^{-1/4}+
\left(
\dfrac{\sqrt{\pi}}{\zeta(3/2)^2}+\dfrac{\zeta(1/2)}{2\sqrt\pi\,\zeta(3/2)}
\right)\tau^{-1/2}
+O(\tau^{-3/4})\nonumber\\
&&\approx
0.5537\,\tau^{-1/4} + 0.102\,\tau^{-1/2}+O(\tau^{-3/4})
\label{nw05frc}
\end{eqnarray}
\end{widetext}
In the following, we shall always take the leading correction to
scaling into account, since  it considerably modifies the behaviour. 

\subsection{The case $\varepsilon>0$}

Let us now analyse the case in which $\varepsilon>0$. We proceed
exactly as in the previous subsection, and we wish to know the
structure of the set $\Sigma_t$. The probability that a particle at
$0$ and another particle at $k$ both belong to $\Sigma_t$, in the case
$k<t$, is still given by the probability, which we now call
$p_\varepsilon(k)$, that a random walk, which starts at the origin and
takes a step to the right, does not return to the origin before time
$k$. The important difference is now that the walk is biased, that is,
that a step to the right now has probability $1/2+\varepsilon$,
whereas a step to the left has probability $1/2-\varepsilon$.

The probability $p_\varepsilon(k)$ has the property that it {\em
  saturates\/} to a positive value $p_\varepsilon(\infty)$ as
$k\to\infty$. More specifically, this saturation happens when $k\sim
k_c(\varepsilon)$, where $k_c(\varepsilon)\sim\varepsilon^{-2}$ as
$\varepsilon\to0$ \cite{Weiss, Hughes}.  When $\varepsilon\to0$ we
have $p_\infty(\varepsilon)\simeq\varepsilon$.  The set $\Sigma_t$ is
therefore a fractal set with a cutoff which is either at $t$ or at
$\varepsilon^{-2}$, depending on which is smaller. The correlation
function in this case is $(kt)^{-1/2}$ for $k\ll\varepsilon^{-2}$,
whereas it goes as $\varepsilon t^{-1/2}$ for
$k\sim\varepsilon^{-2}$. Of course, $k\simeq\varepsilon^{-2}$ and
$k<t$ are only compatible if $t>\varepsilon^{-2}$.

If, on the other hand, $\varepsilon<0$, that is, if we are looking at
the number of surviving particles of the {\em minority\/} species,
then the probability of surviving for $k$ time steps decays
exponentially in $k$ as $k$ becomes larger than a characteristic value
$k_c(\varepsilon)\simeq\varepsilon^{-2}$. This means that there are
essentially no minority particles when $t>k_c(\varepsilon)$, in other
words, after a time of order $\varepsilon^{-2}$. We shall call this
time $t_1(\varepsilon)$. We will again obtain this time scale from the
scaling at the crossover point at the end of this section, with the
implication that the relatively few particles that survive at such
times are very close to each other.

When all minority particles have annihilated, and only the majority
species remains, the surviving particles lie on a fractal of dimension
$1/2$ with a lower cutoff $t_1(\varepsilon)^{1/2}$ and an upper cutoff
$t_1(\varepsilon)$, where $t_1(\varepsilon)$ is the time at which the
minority species disappears. At that time, one has $N(0)^{1/4}$
particles, where $N(0)$ is the initial number of majority
particles. These particles will now undergo diffusive annihilation
with diffusion constant $D_{eff}$ (Appendix \ref{diffconst}), starting
from a fractal distribution. Actually, not only these remaining
majority particles, but all excess particles $c_{ex}(t)$ will undergo
diffusive annihilation on a fractal of fractal dimension $1/2$ from
the beginning, except when $\varepsilon=\pm 1/2$ in which case all
particles are excess particles.

Let us determine the crossover time $t_1(\varepsilon)$. This is the
time at which we cross from the $t^{-3/4}$ initial behaviour, to the
long-time $t^{-1/4}$ behaviour. The latter, as follows from the above,

\begin{equation}
 at_1(\varepsilon)^{-3/4}=b(\varepsilon)t_1(\varepsilon)^{-1/4}
 \label{scrvr}
\end{equation}
where the coefficient $a$ does not depend on $\varepsilon$, as the
decay at the beginning does not depend on the initial
concentration. On the other hand, $b(\varepsilon)\sim \varepsilon $,
as $p_\infty(\varepsilon)\simeq\varepsilon$ for $\varepsilon
\rightarrow 0$.  Simplifying equation (\ref{scrvr}), we get
\begin{equation}
 t_1(\varepsilon) \sim \varepsilon^{-2} 
 \label{t1cr}
\end{equation}
which is compatible with the previous description. For
$\varepsilon\to0$, this crossover time $t_c$ will diverge and hence
will scale with system size $L$ for finite $L$.

For $\varepsilon=0$, the $t^{-1/4}$ decay is difficult to see, as
there remain very few particles at this late stage.  For $\varepsilon
\neq 0$, it is respectively easier to detect the $t^{-1/4}$ (which is
the leading term) decay of concentration on the fractal, because there
remain more particles in this situation.  As the dynamics is diffusive
in this regime, the fractal structure will eventually fade out to the
uniform distribution. Hence there will be yet another crossover and
the concentration will decay as $t^{-1/2}$ at very late time.  If the
crossover time for this second crossover is $t_2(\varepsilon)$ we can
write
 \[
b(\varepsilon)t_2(\varepsilon)^{-1/4}=c\,t_2(\varepsilon)^{-1/2}
 \]
where the coefficient $c$ does not depend on the initial concentration
and hence on $\varepsilon$. This gives
\begin{equation}
 t_2(\varepsilon) \sim \varepsilon^{-4}
 \label{t2cr}
\end{equation}
To detect and measure this second crossover time $t_2(\varepsilon)$ is
quite difficult, as the number of remaining particles is very low and
the time very large.  To be able to observe it, we need to use
comparatively large values of $\varepsilon$, for which other effects,
such as the $-3/4$ initial decay, are not so clearly visible.

\section{Numerical Results and scaling}

\subsection{Results for  $\varepsilon = \pm 1/2$}
\begin{figure}[ht]
\includegraphics[width=6cm,angle=270]{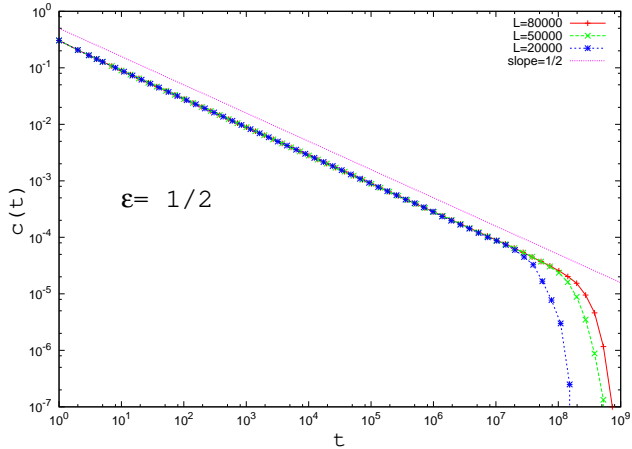}
\caption{(Color Online) The concentration $c(t)$ for the total number of particles decay as $t^{-1/2}$.}
\label{nwpc1}
\end{figure}

For $\varepsilon = \pm 1/2$, there exists only one kind of particles
(either $+v$ or $-v$ velocity particles) and hence the excess number
of particles $c_{ex}(t)=c(0)$. As discussed above, the system is
purely diffusive in this case and, thus, the concentration $c(t)$
decays with time $t$ as $t^{-1/2}$ (Fig. \ref{nwpc1}).

 The domain size of positive or negative velocity particles (depending
 on $\varepsilon = + 1/2$ or $-1/2$) cannot change and is equal to the
 system size from the beginning. The probability distribution of
 inter-particle distances between two neighbouring particles,
 increases linearly with the domain size (due to diffusion) for
 smaller domains \cite{bald} and then drops exponentially, as expected
 for diffusion. The persistence probability decays exponentially with
 time as the walkers are ballistic and wipe out the persistence of all
 the lattice sites.
 
 \subsection{Results for  $\varepsilon = 0$}
\begin{figure}[ht]
\includegraphics[width=6cm,angle=270]{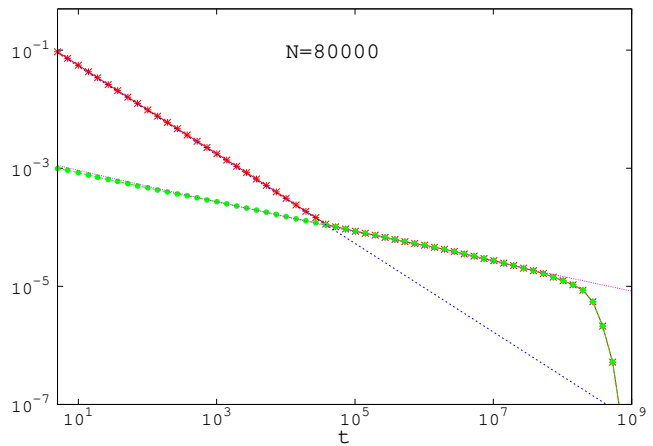}
\caption{(Color Online) The concentration $c(t)$ (red points) for the total number of particles
  decay as $t^{-z_1}$, with $z_1 =3/4$ (blue dotted line) before the crossover and
  following the equation (\ref{nw05frc}) after the crossover. The
  decay of $c_{ex}(t)$ (green points), concentration for the excess number of
  particles is also plotted. $n(t)/\sqrt{N}$, with $\lambda=1.5$ is
  plotted as the theoretical curve (pink dotted line), where the expression of $n(t)$ is
  given by equation (\ref{nwfrct}). The decay of the concentrations
  is plotted starting with initially $N=80000$ particles.}
\label{nw}
\end{figure}

 When $\varepsilon = 0$ the number of particles of each kind is equal
 on average. As discussed above, the system will undergo a crossover
 from ballistic annihilation with superimposed diffusion to purely
 diffusive annihilation on a fractal.  Figure \ref{nw} shows the decay
 of the concentration $c(t)$ for the total number of particles and the
 decay of $c_{ex}(t)$, the concentration of the excess number of
 particles. Both the concentrations are normalized by total number of
 initial particles, denoted by $N$. The concentration $c(t)$ decays as $t^{-3/4}$
 (equation \ref{powerz1}) before the crossover and following equation
 (\ref{nw05frc}) after the crossover (Fig. \ref{nw}).
 
 The initial concentration for the excess particles is
 $c_{ex}(0)=1/ \sqrt{N}$. If the excess particles, which decay due to
 diffusive annihilation, are assumed to lie on a fractal of dimension
 $1/2$ from the beginning, then they will decay according to equation
 (\ref{nw05frc}). The plot of Eq. \ref{nw05frc} with the numerical
 data for $c_{ex}(t)$ (Fig. \ref{nw}) shows excellent agreement,
 supporting the assumption made above.
 
After the crossover, the total number of particles is equal to the
excess number of particles, as there exists only one kind of particles
in this regime.  However the number of particles left after the
crossover is quite limited, ${\mathcal O}(N^{1/4})$ and a very large
number of configurations (more than $2\times10^3$) are needed to
attain the proper statistics.  Further, as previously mentioned, one
has to consider $n(t)/\sqrt{N}$ to fit the decay of the fraction of
excess particles from the beginning.
\begin{figure}[ht]
\includegraphics[width=6cm,angle=270]{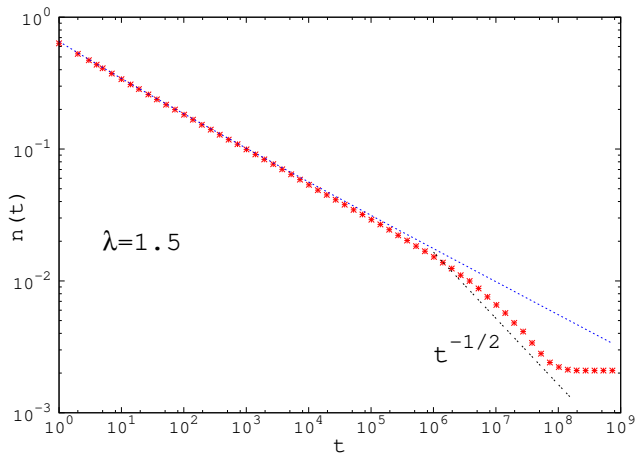}
\caption{(Color Online) $n(t)$, normalized by initial number of particles $n(0)$ for
  each configurations, are plotted as a function of time $t$ (red dots). The blue line is the 
  theoretical plot, which is the plot of equation (\ref{nwfrct}) with
  $\lambda=1.5$ (that is the plot of equation \ref{nw05frc}). In this plot the system size \textbf{ $L=150000$} and the
  average initial number of particles are\textbf{ $\langle
  n(0)\rangle=O(680)$}. }
\label{fnw_fractal}
\end{figure}
Hence to check the expression of equation (\ref{nw05frc}) directly, we
have also studied the dynamics starting from an initial configuration,
where all the particles have the same velocity and are distributed on
a fractal [Fig \ref{fnw_fractal}]. The numerics show a good agreement
with the theoretical plot.

\begin{figure}[ht]
\includegraphics[width=8.6cm,angle=0]{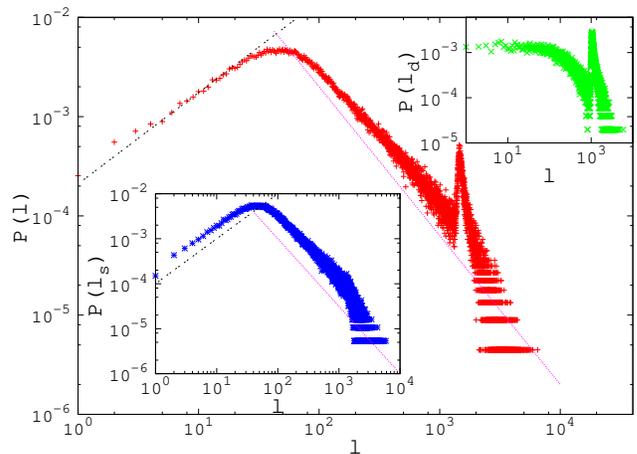}
\caption{(Color Online) The distribution $P(l)$ for the inter-particle distances
  between two neighbouring particles, independent of their velocity is
  plotted in the main plot. Inset at the bottom shows the distribution
  $P_s(l)$, for the same velocity particles and the top right inset
  shows the distribution $P_d(l)$, for two different velocity
  particles. Pink line indicates $l^{-3/2}$ decay for both the main plot and at the bottom inset. 
  Similarly black dashed line shows the linear increase. }
\label{distnn}
\end{figure}

At late times, the distribution of interparticle distances between two
neighbouring particles was also studied.  This distribution for two
same velocity neighbouring particles, which is denoted by $P_s(l)$,
goes as $l^{-3/2}$ for large $l$ [Inset at the bottom of
  fig. \ref{distnn}]. This is evidence that inside a domain of same
velocity particles, the particles are distributed on a fractal of
dimension $1/2$, while $P_s(l) \sim l$, for small $l$.  That this is
due to the diffusion-limited annihilation follows, for example, from
the exact results of \cite{DBA}, where it is shown that the
interparticle distribution function for diffusion-limited annihilation
grows linearly as $x$ for $x$ much less than the average interparticle
distance.  On the other hand we have also computed $P_d(l)$, the
distribution of inter-particle distances between two different
velocity neighbouring particles.  The distribution is almost flat and
then has an exponential decay [Top right inset of figure
  \ref{distnn}].  At the part of the exponential decay the value of
$P_d(l)$ suddenly increases indicating that the probability of having
some large value of $l$ is very high. This is due to the fact that the
two domains or fractals of same velocity particles are moving apart
from each other and the distance increases almost linearly with
time. Indeed, this sudden increase of probability is not observed in
the distribution function $P_s(l)$.
 
$P(l)$, the general distribution function for inter-particle distances
between two neighbouring particles, where the neighbouring particles
can have any velocity, is also computed [The main plot of
  Fig. \ref{distnn}]. As a combined effect of $P_s(l)$ and $P_d(l)$,
$P(l) \sim l$, for small $l$ and also goes as $l^{-3/2}$ for large $l$
[Fig. \ref{distnn}], before the exponential decay.  This indicates
that the fractal structure is dominating at late times, right up to
just before the crossover time, when a small number of minority
particles is still left in the lattice. As an effect of $P_d(l)$, the
sudden increase of probability for a large value of $l$, is also
present in this general distribution function $P(l)$.

\begin{figure}[ht]
\includegraphics[width=8.6cm,angle=0]{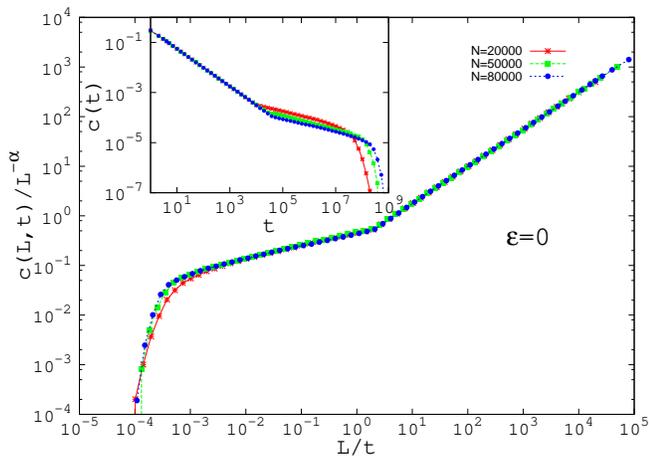}
\caption{(Color Online) The collapsed plot of scaled data of concentration of
  particles with $\alpha=0.75$ for different number of particles for
  $\varepsilon=0$.  Inset shows the raw data. The collapse is not good
  for the exponentially decaying part, as the scaling theory 
  applies to the two power law regions only. The average number of
  particles is also considerably less than $one$ in this
  exponentially decaying region. }
\label{nw_colp}
\end{figure}

The decay of $c(t)$, the concentration of total number of particles,
has a dependence on the initial number of particles [inset of
  Fig. \ref{nw_colp}]. After the crossover, when $c(t,L)$ is equal to
$c_{ex}(L,t)$. Finite-size scaling analysis can be done using the
scaling form
\begin{equation}
  c(t,L) \sim L^{-\alpha} f(L/t)
 \label{fss_epsi0}
 \end{equation}
where $f(x) \rightarrow x^{\alpha}$ with $\alpha=3/4$, for $x
\rightarrow \infty$ and $f(x) \rightarrow x^{-1/4}$ when $x \ll1$.
The raw data as well as the scaled data using $\alpha=3/4$ are shown
in Fig. \ref{nw_colp}.

The domain size for same velocity particles is defined as the number
of consecutive same velocity particles (either $+v$ or $-v$). $S_d(t)$
is the average over all the domains of same velocity particles at the
time $t$. On the other hand, lattice domain size is defined by the
number of lattice sites occupied by these domain of same velocity
particles and $S_{ld}(t)$ is the average over all these lattice
domains at the time $t$. Both the average domain sizes $S_d(t)$ and $S_{ld}(t)$ are 
normalised by the size of the lattice. The average value of the lattice domain size can 
not go beyond $0.5$ as the entire lattice can be occupied by either the positive or the 
negative velocity particles (with $50\%$ probability) at very late time.
\begin{figure}[ht]
\includegraphics[width=8.6cm,angle=0]{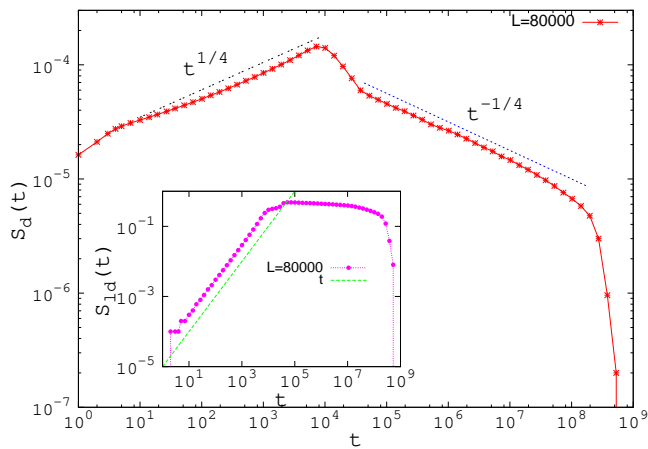}
\caption{(Color Online) The plot for change of $S_d(t)$, the average domain size (normalised by the size of the lattice) for
  same velocity particles, with time $t$. Inset shows the change of
  lattice domain size $S_{ld}(t)$  (normalised by the size of the lattice) for the same velocity particles,
  with time $t$.}
\label{dsz}
\end{figure}

At some late time, but before the crossover, that is, in the region
where $c(t) \sim t^{-3/4}$, the average interparticle distance
increases as $t^{3/4}$. Now $S_{ld}(t)$, the average lattice domain
size increases linearly in time, due to the ballistic nature of the
particles [Inset of Fig.  \ref{dsz}]. Hence $S_d(t)$, the average
domain size for same velocity particles increases with time as
$t/t^{3/4}=t^{1/4}$, in this region [Fig. \ref{dsz}]. After the
crossover, there exists only one domain and hence $S_{ld}(t)$ become
constant. In this region after the crossover, $S_d(t) \sim t^{-1/4}$
as the concentration of particles decreases as $t^{-1/4}$
[Fig. \ref{dsz}].  We have also computed the average interparticle
distance between two different velocity neighbouring particles (not
shown), which should increase linearly with time. This quantity
increases with time with an effective exponent $0.93$ in our
simulation, which is a little ambiguous but agrees with the
observation made in \cite{baldif}.

\begin{figure}[ht]
\includegraphics[width=6cm,angle=270]{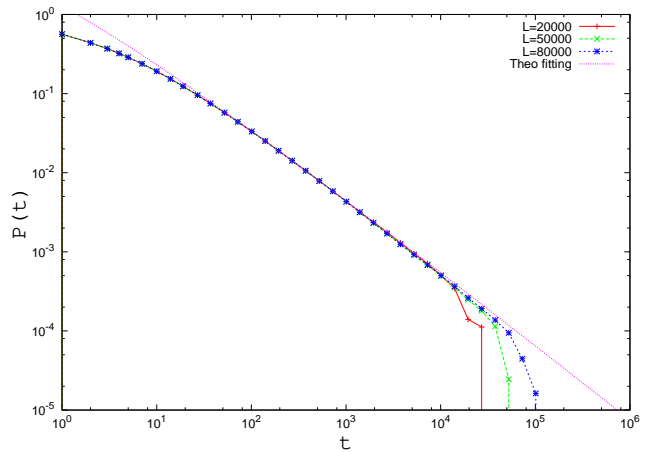}
\caption{(Color Online) Decay of the persistence probability $P(t)$ with time $t$ for
  different system sizes. The theoretical fitting curve is plotted
  following equation \ref{perfit} where the fitting parameters are
  $a=0.45$ and $b=1.27$.}
\label{per}
\end{figure}

The persistence probability does not show any finite size dependence
before the crossover and fits quite well to the form [Fig. \ref{per}]
\begin{equation}
P(t) \simeq a\dfrac{\log(t)}{t} +\frac{b}{t}
 \label{perfit}
\end{equation}
with $a=0.45 \pm 0.01$ and $b=1.27 \pm 0.05$, obtained by least square
fitting of the numerical data.  Although the fit certainly is quite
good and the functional dependence remarkably simple, we would like to
mention that we have no rationale at all for this behaviour.
After the crossover, the persistence probability decays exponentially
as the remaining same velocity ballistic particles wipe out the
persistence of all the remaining sites.

\subsection{Results for $0 < \varepsilon < 1/2$}

\begin{figure}[ht]
\includegraphics[width=8.6cm,angle=0]{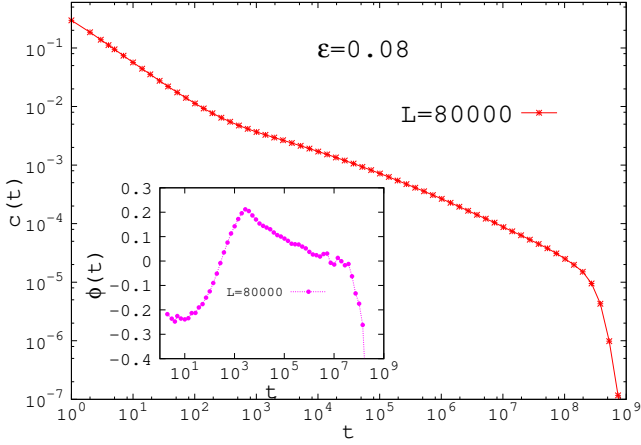}
\caption{(Color Online) Decay of $c(t)$, the concentration for the total number of
  particles, with time $t$ is plotted in the main plot. Inset shows
  the change of $\phi(t)$ with time $t$.}
\label{numder08}
\end{figure}

In this section we will present the numerical results for the cases in
which the number of positive velocity and negative velocity particles
are not equal on average. For $\varepsilon \neq 0$, as previously
mentioned, there exist three different dynamical regimes where the
concentration decays with different exponent values. However, due to
the finiteness of the system, the numerics is challenging because the
regimes are not that well separated and thus, the crossover times are
not very clean, as can be seen in Fig. \ref{numder08} where we plot
the behavior of $c(t)$. 

To better analyze the behavior of the concentration, we have
computed the function $\phi(t)$ defined as
 \begin{equation}
  \phi(t) = \dfrac{d}{d(\log(t))}[\log (\sqrt{t}~ c(t))].
  \label{derv}
 \end{equation}
Thus, when $c(t) \sim t^{-3/4}$, $\phi(t)=-1/4$ and so on. In particular,
when $c(t) \approx~ at^{-3/4}+bt^{-1/4}$, $\phi(t)$
changes from $-1/4$ to $1/4$, whereas when $c(t) \approx~
bt^{-1/4}+ct^{-1/2}$, $\phi(t)$ changes from $1/4$ to $0$. The inset of
figure \ref{numder08} shows the change of $\phi(t)$ with time $t$ for
$\varepsilon =0.08$.
 
In this case $c_{ex}(t)$, the concentration for the excess number of
particles, decays as $t^{-1/4}$ initially and then as $t^{-1/2}$
at late times.
 
The concentration of the total number of particles is a function of
$\varepsilon$ also, so we write it explicitly as $c(L,\varepsilon,t)$.
We now turn to the scaling behavior of $c(L,\varepsilon,t)$. As the
first crossover time is $t_1(\varepsilon) \sim \varepsilon^{-2}$ (Eq.
\ref{t1cr}), the dimensionless quantity controlling the crossover
between the first two dynamical regimes will be
$\varepsilon^2t$. Similarly the dimensionless quantity for the
crossover between the second and third regimes will be
$\varepsilon^4t$, as the second crossover time is $t_2(\varepsilon)
\sim \varepsilon^{-4}$ (Eq. \ref{t2cr}).

\begin{figure}[ht]
\includegraphics[width=6cm,angle=270]{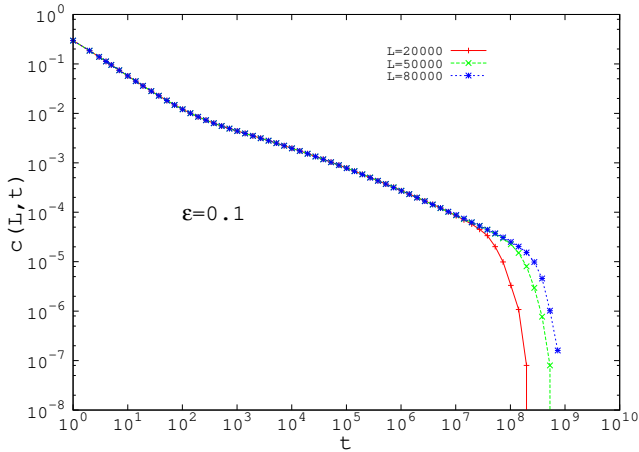}
\caption{(Color Online) Decay of $c(t)$, for different finite sizes, starting from
  $c(0)=1$. Decay of $c(t)$ is size independent for any $\varepsilon
  \neq 0$. The plots are for $\varepsilon=0.1$ here.}
\label{nosize}
\end{figure}

Unlike the case of $\varepsilon = 0$, in this case 
we are considering infinite systems, since finite size effects would only be significant when 
$\varepsilon\ll1$. Hence the
first scaling ansatz is that, except for very long times,
\begin{equation}
 c(L,\varepsilon,t) = c(\varepsilon,t),\qquad\hbox{\rm for all }\varepsilon \neq 0.
\end{equation}
That is, there will be no system size dependence of $c(t)$ for a
constant $\varepsilon \neq0$ (for all $\varepsilon$), which is indeed
borne out by the simulations [Fig. \ref{nosize}], except for the long
time exponential decay where this scaling ansatz does not hold.

\begin{figure}[ht]
\includegraphics[width=8.6cm,angle=0]{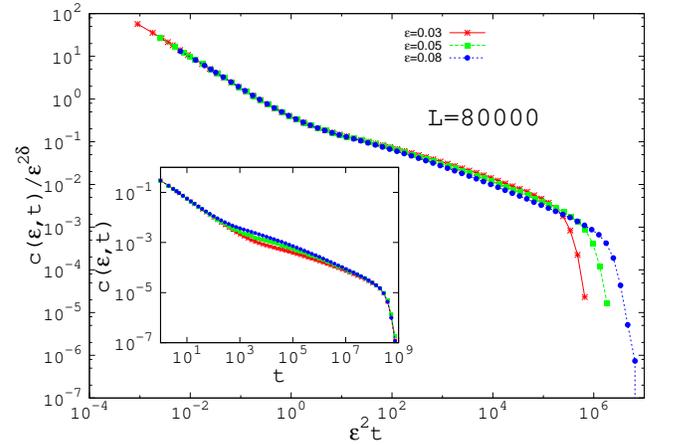}
\caption{(Color Online) The collapsed plot of scaled data of concentration of
  particles with $\delta=0.75$ for different $\varepsilon \neq 0$.
  Inset shows the raw data. The scaled data gives a good collapse for
  first two dynamical region where the exponent values are $3/4$ and
  $1/4$ respectively.   
}
\label{nw_1_colp}
\end{figure}

Now we will discuss the scaling laws involving $\varepsilon$ and
$t$, which hold for $|\varepsilon| \ll 1$.  The scaling function
describing the first two dynamical regimes can be written as
\begin{equation}
 c(\varepsilon,t) \sim \varepsilon^{2\delta} f(\varepsilon^2t) ,\qquad\hbox{\rm for }\varepsilon \ll 1
 \label{fss1_epsi}
\end{equation}
where $f(x) \rightarrow x^{-\delta}$ with $\delta=3/4 $, for $x \ll 1$
and $f(x) \rightarrow x^{-1/4}$ when $x \gg1$.  The raw data as well
as the scaled data using $\delta=3/4$ are shown in
Fig. \ref{nw_1_colp}. The collapse is good for first two regimes.

The scaling analysis for the second and third dynamical regime (where
$\varepsilon^4t$ is the relevant dimensionless quantity) can be
carried out using the scaling form
\begin{equation}
 c(\varepsilon,t) \sim \varepsilon^{4\eta} g(\varepsilon^4t),\qquad\hbox{\rm for all }\varepsilon \ll 1
 \label{fss2_epsi}
\end{equation}
where $g(x) \rightarrow x^{-\eta}$ with $\eta=1/2 $, for $x
\rightarrow \infty$ and $g(x) \rightarrow x^{-1/4}$ when $x \ll1$.
\begin{figure}[ht]
\includegraphics[width=8.6cm,angle=0]{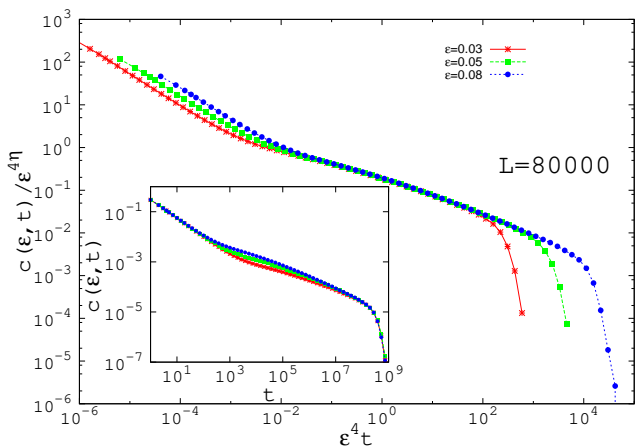}
\caption{(Color Online) The collapsed plot of scaled data for the concentration of
  particles with $\eta=1/2 $ for different $\varepsilon \neq 0$.
  Inset shows the raw data. The scaled data give a good collapse for
  second and third dynamical region where the exponent values are
  $1/4$ and $1/2$ respectively. Of course, data collapse does not
  occur for the exponentially decaying part, as the scaling theory
  applies for the power law regions only.  The average number of
  particles is again considerably less than $one$ in this
  exponentially decaying region. }
\label{nw_2_colp}
\end{figure}
The behaviour of $c(\varepsilon,t)$ for $ \varepsilon^2 t\gg 1$ in
equation (\ref{fss1_epsi}), is the same as that of equation
(\ref{fss2_epsi}) for $\varepsilon^4 t\ll 1$ (in both cases one gets
that $c(\varepsilon,t)\sim \varepsilon t^{-1/4}$), so both scaling
laws are consistent with each other.

The raw data as well as the scaled data using $\eta=1/2 $ are shown in
Fig. \ref{nw_2_colp}. The collapse is good for second and third
dynamical regimes.

$S_{ld}(t)$, the average lattice domain size does not increase
linearly in time but increases faster than that (though it does not
appear to grow as a well defined power of time $t$) and then
saturates. $S_d(t)$, the average domain size for same velocity
particles again increases with time initially and then
follows the concentration $c(\varepsilon,t)$ at late times when
minority particles do not exist anymore.

$P(t)$, the persistence probability decays exponentially for all
values of $ \varepsilon \neq 0$. This is due to the presence of more
particles of one velocity than of the other; the majority
particles wipe out the persistence of all the sites very quickly.

\section{Conclusion}
To summarise, we have considered the model of annihilating particles
moving ballistically with a superimposed diffusion.  As shown in
\cite{ballistic,baldif}, the particle concentration was found to
decay as $t^{-3/4}$, that is, faster than either the purely ballistic
or the purely diffusive case, both of which decay as $t^{-1/2}$. This
result, however, fails once all particles are of the same species,
that is, either right- or left-going. If the initial condition has an
equal concentration of the two kinds of particles, there will
nevertheless remain a number $N^{1/4}$ of particles of one velocity
after all the particles of opposite velocity are annihilated. This
follows from the fact, which we confirmed numerically, that the number
of excess particles decays as $t^{-1/4}$. Since this number starts out
at $\sqrt N$, it decays as $\sqrt N~t^{-1/4}$, which becomes equal to
$t^{-3/4}$ when $t\sim N$. From this, the fact that the number of
particles remaining at this time, scales as $N^{1/4}$, follows
immediately.

Thus the number of particles remaining after one kind of particle has
been eliminated goes to infinity as $N$ does. It therefore makes sense
to ask: with what power of $t$ do the remaining particles decay? We
have shown that they decay with the exponent $t^{-1/4}$ and have
provided a rationale for this behaviour in terms of Alemany's result,
that annihilating particles which start out distributed on a fractal
of dimension $d_f$, decay as $t^{-d_f/2}$. Since it can be argued that
the excess particles are constrained to lie on a fractal of dimension
$d_f=1/2$, the result readily follows.

Finally, we have also looked at the case in which the initial
concentrations of left- and right-going particles differ, their
initial values being given by $1/2-\varepsilon$ and $1/2+\varepsilon$
respectively.  If $\varepsilon\ll1$, we have shown that two crossovers
arise: one from the usual $t^{-3/4}$ behaviour to the $t^{-1/4}$
behaviour described in the preceding paragraph. This crossover arises
at a time $t_(\varepsilon)\sim\varepsilon^{-2}$. A second crossover to
an ordinary $t^{-1/2}$ decay, characteristic of ordinary
diffusion-limited annihilation in one-dimension, is observed at a
crossover time $t_2(\varepsilon)\sim\varepsilon^{-4}$. This second
crossover is not observed for $\varepsilon=0$, since the number of
remaining particles for that regime turns out to tend to zero as
$N\to\infty$.

\section{Acknowledgement}
Financial support from CONACyT through Projects No. 154586 and Program UNAM - DGAPA PAPIIT IN114014 is acknowledged. S.B acknowledges the support from
DGAPA/UNAM postdoctoral fellowship. The authors also acknowledge the Miztli supercomputer of DGTIC, UNAM for the computational resources under the 
project number SC15-1-S-20. 

\appendix
\renewcommand{\theequation}{A-\arabic{equation}}
\setcounter{equation}{0}
\section{Diffusion constant}
\label{diffconst}

When all the particles are of same velocity (say $+v$), the only
source of diffusion is the random update rule. So whenever a particle
is randomly chosen it will move forward a single lattice site and one
Monte Carlo update will be over. After $N$ such update one Monte Carlo
time step will be over, if there are $N$ particles in the lattice.

We calculate the relative diffusion constant $D_{eff}$ between
two neighbouring particles.  Let $P(l,n)$ be the probability that the
relative distance between two chosen neighbouring particles be $l$
after the $n$th Monte Carlo update, where $ 0 < n < N $. At the
$(n+1)$th update, any one of the two particles can be chosen with a
probability $1/N$, in which case the distance $l$ will be increased or
decreased by one lattice site; or neither of these particles will be choosen
with a probability $(1-2/N)$, in which case the distance between
the particles remains unchanged. Thus we can write the following equation
\begin{eqnarray}
 P(l,n+1)&=&\dfrac{1}{N} 
 \bigg[
 P(l-1,n) +  P(l+1,n)
\bigg] \nonumber\\
&& + 
 \left(
 1-\dfrac{2}{N}
 \right)
  P(l,n)
\label{apmaster}
\end{eqnarray}
Taking the Fourier transform of (\ref{apmaster}) we get 
\begin{equation}
 \hat{P}(\omega,n+1)=
 \left(
 1-\dfrac{2}{N}+\dfrac{2}{N}\cos{\omega}
 \right)
 \hat{P}(\omega,n)
 \label{ftmaster}
\end{equation}
where $\hat{P}(\omega,n)$ is the Fourier transform of $P(l,n)$.

We have to repeat this update $N$ times to complete one Monte Carlo
time step. Initially, for $n=0$,
\[
\hat{P}(\omega,n)=\hat{P}(\omega,0)=\exp(i\omega \ell)
\]
where $\ell$ is the initial distance between these two
particles. After $N$ update, equation \ref{ftmaster} becomes
\begin{equation}
 \hat{P}(\omega,N)=
 \left(
 1-\dfrac{2}{N}+\dfrac{2}{N}\cos{\omega}
 \right)^N\hat{P}(\omega,0)
 \label{Nupdate}
\end{equation}
After $t$ such time steps equation \ref{Nupdate} becomes 
\begin{equation}
 \hat{P}(\omega,t)=
 \left(
 1-\dfrac{2}{N}+\dfrac{2}{N}\cos{\omega}
 \right)^{Nt}\exp(i\omega \ell)
 \label{ttime}
\end{equation}
Doing a Taylor series expansion of cosine and exponential function in (\ref{ttime})
 \begin{eqnarray}
\hat{P}(\omega,t)&=&
  \left(
  1-\dfrac{1}{N}\omega^2+\ldots
  \right)^{Nt}\!
  \left(
  1+i\omega \ell -\dfrac{\omega^2 \ell^2}{2}+\ldots
  \right)
\nonumber\\
&=&1+i\omega \ell -\dfrac{1}{2}(2t+\ell^2)\omega^2+..
\label{fnlft}
 \end{eqnarray}
From which we get 
\[
 \langle
 l
 \rangle=\ell\qquad \hbox{and}\qquad
 \langle
 l^2
 \rangle=2t+\ell^2,
\]
which implies that 
\begin{equation}
\langle l^2\rangle -\langle l \rangle^2=2D_{eff}~t=2t.
\end{equation}
This in turn gives 
\begin{equation}
D_{eff}=1.
\end{equation}

\renewcommand{\theequation}{B-\arabic{equation}}
\setcounter{equation}{0}
\section{Decay of concentration starting from initial fractal distribution of particles}
\label{fractal}
Let $F (x, t)$ be the probability at time $t$ in which a random walker, starting at 
$x = 0$, reaches site $x$ for the first time. For any lattice with 
translational invariance, the Laplace transform of $F (x, t)$ will be given 
by $\tilde{F} (x, u)$ \cite{Weiss,Hughes}. It can be written as
\[
 \textit{L}_u\{{F (x, t)\}}=\tilde{F} (x, u) = \tilde{\xi}(u)^{-x}~,
        {\rm  with}
\]
\begin{equation}
  \tilde{\xi}(u)^{-x}=1 + (u/D) + \sqrt{2(u/D) + (u/D)^2}
 \label{c1}
\end{equation}
where $D=D_{eff}$ is the relative diffusion constant.  If
$\tilde{P}(v)$ is the Laplace transform of the probability
distribution function $P(x)$, for initial inter-particle distances for
two nearest particles then the number of particles $n(t)$
at time $t$ (normalized by initial number of particles $n(0)$) will be
\cite{alemany}
\begin{equation}
 n(t)=\textit{L}_u^{-1}
 \left\{
u^{-1} \dfrac{1-\tilde{P}(v=\ln\tilde{\xi}(u/2))}{1+\tilde{P}(v=\ln\tilde{\xi}(u/2)) }
 \right\} 
 \label{cnt}
\end{equation}
If initially the particles are distributed on a fractal with fractal
dimension $d_f$ ($0<d_f<1$), then we can write
\begin{equation}
 P(x)=\dfrac{1}{\zeta(\lambda)}\sum_{l=1}^{\infty} l^{-\lambda}\delta(x-l)
\label{cindist}
\end{equation}
where $\lambda = d_f +1$ and hence $1<\lambda<2$. 
The Laplace transform of equation (\ref{cindist}) will be 
\begin{eqnarray}
 \tilde{P}(v)&=&\int_0^{\infty}e^{-xv}P(x)\,dx\nonumber\\
 &=&\dfrac{1}{\zeta(\lambda)}\sum_{l=1}^{\infty} l^{-\lambda}e^{-lv}\nonumber\\
&=&\frac1{\zeta(\lambda)}
\Phi\left(
e^{-v}, \lambda, 1
\right)
 \label{c2}
\end{eqnarray}
where $\Phi(z, s, a)$ is defined in \cite{erd}, Section 1.11. 
Thus we can write, following equation (8) of the same Section of \cite{erd}:
\begin{eqnarray}
  \tilde{P}(v)&=&\dfrac{1}{\zeta(\lambda)}
  \left[
  \Gamma(1-\lambda)v^{\lambda-1}+\sum_{k=0}^{\infty}\dfrac{\zeta(\lambda-k)}{k!}(-v)^k
  \right]
  \nonumber\\
  &\simeq& 1+\dfrac{\Gamma(1-\lambda)}{\zeta(\lambda)}v^{\lambda-1} - \dfrac{\zeta(\lambda-1)}{\zeta(\lambda)}v +\ldots
\label{cpv}
\end{eqnarray}
Note that this function has no special name in \cite{erd}. It is, however, related to functions denoted as {\em polylogarithms},
treated, for example, in \cite{wood}. 
Now we have the elements to evaluate $n(t)$ given in 
(\ref{cnt}). First we write 
\begin{equation}
 v=\ln\tilde{\xi}(u/2))\approx (u/D)^{1/2}-\frac{3}{8}(u/D)^{3/2}+...
\end{equation}
and hence, to leading order, we have
 \begin{eqnarray}
v^{\lambda-1}&=&(u/D)^{(\lambda-1)/2}
\left[
1+O\left(
u/D
\right)
\right]
\label{cv} 
 \end{eqnarray}
Then inserting $\tilde{P}(v)$ given by equation (\ref{cpv}), along 
with equation (\ref{cv}), in equation (\ref{cnt}), we get 
\begin{widetext}
\begin{equation}
n(t)=\textit{L}_u^{-1}
\left\{
u^{-1}
\dfrac{-\Gamma(1-\lambda)
(u/D)^{(\lambda-1)/2}+
\zeta(\lambda-1)
(u/D)^{1/2}}
{2\zeta(\lambda)+\Gamma(1-\lambda) 
(u/D)^{(\lambda-1)/2}
-\zeta(\lambda-1)(u/D)^{1/2}}
\right\}.
\end{equation}
\end{widetext}
This implies
 \begin{widetext}
\[
 n(t)\approx\textit{L}_u^{-1}
 \left\{
-\dfrac{\Gamma(1-\lambda)}{2\zeta(\lambda)}
\dfrac{u^{(\lambda-3)/2} }{D^{(\lambda-1)/2}}
+\dfrac{\Gamma(1-\lambda)^2}{4\zeta(\lambda)^2}
\dfrac{u^{\lambda-2} }{D^{\lambda-1}}+ 
\dfrac{\zeta(\lambda-1)}{2\zeta(\lambda)}\frac{u^{-1/2}}{D^{1/2}}
 + O\left(
 {D^{-\lambda/2}}u^{\lambda/2-1}
 \right)
 \right\},
\] 
 \end{widetext}

which in turn leads to:
\begin{eqnarray}
n(t)&=&a_1\tau^{-(\lambda -1)/2} +
a_2\tau^{-(\lambda-1)}\nonumber\\ 
&&+a_3\tau^{-1/2} +O\left(
\tau^{-\lambda/2}
\right),
\end{eqnarray}
where  $\tau=Dt$ and
\begin{subequations}
\begin{eqnarray}
a_1&=&- \dfrac{\Gamma(1-\lambda)}{2\zeta(\lambda)\Gamma(\frac{3-\lambda}{2})}
\\
a_2&=&\dfrac{\Gamma(1-\lambda)}{4(1-\lambda) \zeta(\lambda)^2}\\
a_3&=& \dfrac{\zeta(\lambda -1)}{2\zeta(\lambda)\sqrt{\pi}}.
\end{eqnarray}
\end{subequations}

\end{document}